\documentclass[twocolumn,showpacs,preprintnumbers,amsmath,amssymb,prb]{revtex4}

\usepackage{graphicx}% Include figure files
\usepackage{dcolumn}% Align table columns on decimal point
\usepackage{bm}% bold math

\begin{document}

\title{Perturbation study of nonequilibrium quasi-particle spectra in an
infinite-dimensional Hubbard lattice}

\author{R. J. Heary and J. E. Han}
\affiliation{
Department of Physics, State University of New York at Buffalo, Buffalo, NY 14260, USA}

\date{\today}
\begin{abstract}
A model for nonequilibrium dynamical mean-field theory is constructed
for the infinite dimensional Hubbard lattice.  We impose nonequilibrium
by expressing the physical orbital as a superposition of a left-($L$)
moving and right-($R$) moving electronic state with the respective
chemical potential $\mu_L$ and $\mu_R$. Using the second-order iterative
perturbation theory we calculate the quasi-particle properties as a
function of the chemical potential bias between the $L$ and $R$ movers,
i.e. $\Phi = \mu_L - \mu_R$. The evolution of the nonequilibrium
quasi-particle spectrum is mapped out as a function of the bias and
temperature.  The quasi-particle states with the renormalized Fermi
energy scale $\varepsilon^0_{QP}$ disappear at
$\Phi\sim\varepsilon^0_{QP}$ in the low temperature limit.  The
second-order perturbation theory predicts that in the vicinity of the
Mott-insulator transition at the Coulomb parameter $U=U_c$, there exist
another critical Coulomb parameter $U_d$ ($<U_c$) such that, for
$U_d<U<Uc$, quasi-particle states are destroyed discontinuously when
$(\varepsilon^0_{QP})^2\sim a(\pi k_BT_c)^2+ b\Phi_c^2$ with the
critical temperature $T_c$ and the critical bias $\Phi_c$.
\end{abstract}


\maketitle

In recent years significant experimental progress has been made in the
fabrication of sophisticated electronic heterostructures utilizing
strongly correlated electronic materials. These systems have given
rise to the discovery of rich novel phenomena.  These include: ballistic
transport of electrons through heterostructures of
superconductors~\cite{Bastian}, ferromagnets~\cite{Salafranca}, magnetic
tunneling junctions~\cite{Teresa, Mccartney} and
oxides~\cite{Thiel,Chakhalian,Brinkman,Reyren}.  With this vast array of
strongly correlated heterostructures it is very important to
theoretically understand how the strongly correlated materials in the
bulk limit will behave under nonequilibrium conditions.  In this work we
formulate the nonequilibrium problem in the lattice and present
significant progress towards a more complete understanding of strongly
correlated lattices out of equilibrium.

We are interested in heterostructures of strongly correlated materials
under a finite source-drain bias where the voltage drop occurs mainly at
the interface of the strongly correlated material and the source/drain
leads~\cite{satoshi1,satoshi2}.  Inside the strongly correlated material
the transport is ballistic and driven by the momentum distribution of
imbalanced chemical potentials.  Our focus here is to investigate the
evolution of strongly correlated quasi-particle spectra due to
nonequilibrium driven by an imbalance in the chemical potentials.  In
this work we gain qualitative understanding on how the enhanced
dephasing by the new particle-hole decay channel due to nonequilibrium
modifies the spectral properties.

One of the most practical and powerful theoretical techniques for
studying strongly correlated lattices is the dynamical mean-field theory
(DMFT)~\cite{GeorgesDMFT}.  Within this approach the self-energy of the
strongly correlated lattice becomes momentum-independent and as a result
the problem is reduced to solving a self-consistent interacting impurity
model.  This inherent simplicity of DMFT  makes it a very attractive
tool, especially with the large quantity of exact and perturbative
impurity solvers such as the Hirsch-Fye quantum Monte Carlo
method~\cite{hirschfye}, the numerical renormalization group~\cite{NRG},
and the non-crossing approximation~\cite{NCA1, NCA2}.

As an extension of DMFT to nonequilibrium situations, Okamoto has
theoretically studied the nonlinear transport and spectral properties of
metal-Mott insulator-metal heterostructures where a bias voltage is
applied across the Mott insulator~\cite{satoshi1, satoshi2} using
layered-DMFT~\cite{satoshimillis}.  There he combined the layered-DMFT
technique with the Keldysh Green function approach and invoked the
non-crossing approximation as the impurity solver.  Another system
recently studied is the DMFT limit of the Hubbard model with the
nonequilibrium driven by a uniform high electric
field~\cite{freerickshub}. Others have looked at the DMFT limit of the
Falicov-Kimball model in the presence of a uniform time-dependent
electric field and examined the transient current, quenching of the
Bloch oscillations, and evolution of the spectral
function~\cite{freericksneqdmft, freericksbloch}.  

In our analysis of the problem we choose to start from the metallic
state and analyze how the Fermi liquid is renormalized and eventually
destroyed with multiple chemical potentials. We find that the
quasi-particles strongly depend upon the strength of the chemical
potential difference $\Phi$. At zero $\Phi$ it has long been known that
the system undergoes a metal-insulator transition at
$U=U_c$~\cite{GeorgesDMFT}. The second order iterative perturbation
theory (IPT) approximation gives $U_c=U_{c2}\simeq3.3D$ at zero
temperature, where $D$ is the half bandwidth. As the temperature is
raised the transition remains distinct down to $U_{c1}\simeq 2.6D$.  In
nonequilibrium, the second order IPT shows that the quasi-particles are
destroyed abruptly by the chemical potential bias in the region
$U_d<U<U_c$, where $U_d\simeq 2.3D$. When $U<U_d$ the system exhibits a
smooth crossover from a system of well-defined quasi-particle states to
that without quasi-particle excitations.

\begin{figure}[h]
\includegraphics[angle=0, width=5.cm]{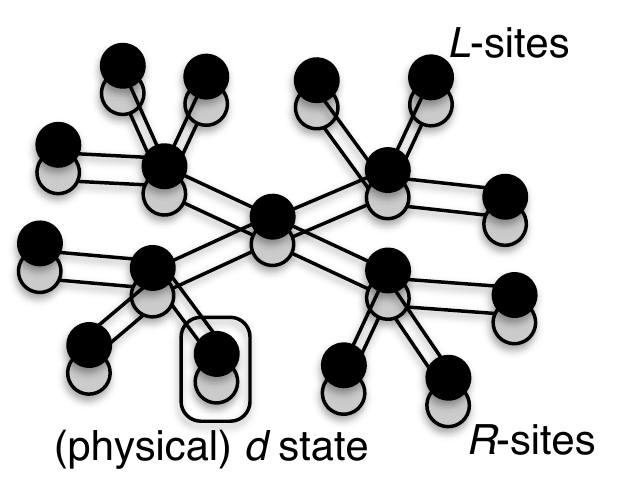}
\caption{The Bethe lattice where 
the (physical) $d$-state is given as a superposition of a left-($L$) and 
a right-($R$) moving state with the 
corresponding chemical potentials $\mu_L = \Phi/2$ and
$\mu_R = -\Phi/2$.  }
\label{fig1}
\end{figure}

We start with a non-interacting $d$-dimensional tight-binding Bethe
lattice.  The tight-binding solutions are grouped evenly into left-
($L$) and right- ($R$) movers with chemical potentials $\mu_L=\Phi/2$
and $\mu_R=-\Phi/2$, respectively. Since we do not study transport
properties in this work, it is not important how the physical site is
decomposed into $L,R$ movers.  The original site orbital
$d^\dagger_{i\sigma}$ at site $i$ with spin $\sigma$ is written as a
superposition of $L$ and $R$ orbitals $d^\dagger_{L/R,i\sigma}$
according to 
\begin{equation}
d^\dagger_{i\sigma}=\frac{1}{\sqrt2}(d^\dagger_{L,i\sigma}+d^\dagger_{R,i\sigma}).
\end{equation}
The interaction terms and any observable quantities ought to
be given in the physical basis $d^\dagger_{i\sigma}$.  The
nonequilibrium DMFT lattice is depicted in Figure~\ref{fig1}. 

The model with which we focus is the DMFT limit ($d\rightarrow\infty$) 
of the Hubbard model $\hat{\mathcal{H}} = \hat{\mathcal{H}}_0 +
\hat{\mathcal{V}}$,
\begin{eqnarray} \hat{\mathcal{H}}_0 &=&
-t\sum_{<i,j>}\sum_{\alpha=L,R;\sigma}\left(d^\dag_{\alpha, i\sigma}
d^{}_{\alpha, j\sigma} + \mathrm{h.c.}  \right) \\ \hat{\mathcal{V}}  &=&
U\sum_i\left(d^\dag_{i\uparrow}d^{ }_{i\uparrow}-\frac{1}{2}\right)
\left(d^\dag_{i\downarrow}d^{ }_{i\downarrow}-\frac{1}{2}\right) .
\end{eqnarray}  
The non-interacting tight-binding part of the Hamiltonian
$\hat{\mathcal{H}}_0$ takes into account the hopping of electrons
between nearest neighbor lattice sites of the Bethe lattice.
The interaction term of the Hamiltonian $\hat{\mathcal{V}}$ is the
particle-hole symmetric on-site Coulomb interaction of strength $U$. 
Both the non-interacting and interacting terms of the Hamiltonian are
homogeneous in real space, which is necessary for our application of
single-site DMFT.
The Hamiltonian may be further simplified by transforming the $L$ and $R$ basis
into the even ($E$) and odd ($O$) superposition of the basis. The creation
operators for the even and odd electron orbitals are 
$d^\dag_{E/O,i\sigma} = (d^\dag_{L,i\sigma} \pm
d^\dag_{R,i\sigma})/\sqrt{2}$. The interaction is only given by the even
(physical) basis and the non-interacting part is decoupled into the
even-odd parts as
$\hat{\mathcal{H}}_0 =
-t\sum_{<i,j>}\sum_{\alpha=E,I;\sigma}(d^\dag_{\alpha, i\sigma}
d^{}_{\alpha, j\sigma} + \mathrm{h.c.}  )$.
Therefore, the time-evolution of even (physical) and odd states
are completely decoupled and the perturbation theory of DMFT is
applied within the physical basis where the nonequilibrium statistics
are imposed on the non-interacting Green functions.

This Hamiltonian is solved using the iterative perturbation
theory~\cite{GeorgesDMFT}. The IPT has been extensively utilized and
well-established in equilibrium DMFT and we anticipate that the second
order IPT gives a qualitatively reasonable description of the
quasi-particle destruction in nonequilibrium. However we caution that
the second order perturbation theory may not capture the exact nature of
the quasi-particles in nonequilibrium. For example, it is known that
second-order perturbation theory does not properly describe the
splitting of the Kondo peak in quantum dot devices with a finite voltage
bias, as seen in the fourth-order perturbation theory~\cite{4thorderPT}
and in nonperturbative solvers~\cite{hanheary,anders}. Therefore, it
still remains to be seen through nonperturbative calculations which of
the following results will hold up.

Now we outline the DMFT self-consistent routine.  The calculation is performed 
in real-time as opposed to real-frequency because close to the transition of a 
system with well defined quasi-particle states to a system with none the bath 
spectral function becomes sharply peaked at the Fermi energy~\cite{Zhang}, 
thus making the real-frequency calculation much more difficult.  
A typical time bandwidth we use is $\sim 3000$.  In the
non-interacting limit the (even) bath green function is given by
\begin{equation}\label{bathGF}
   \left[\begin{matrix} % or pmatrix or bmatrix or Bmatrix or ...
      g^<(t) \\
      g^>(t) \\
   \end{matrix}\right] = \frac{i}{2}\sum_{\alpha=L,R}\int d\epsilon D(\epsilon)
      \left[\begin{matrix} % or pmatrix or bmatrix or Bmatrix or ...
         f_\alpha(\epsilon) \\
         f_\alpha(\epsilon)-1 \\
      \end{matrix}\right]e^{-i\epsilon t} ,
\end{equation}
where $f_L(\epsilon) = f(\epsilon - \frac{\Phi}{2})$ and $f_R(\epsilon)
= f(\epsilon + \frac{\Phi}{2})$ are the Fermi functions for the $L$ and $R$ 
movers.  $D(\epsilon)$ is the non-interacting density of states
(DOS), which we have taken to be semi-circular $D(\epsilon)=\frac{2}{\pi
D^2}\sqrt{D^2-\epsilon^2}$. In the following, $D=1$ is used as the unit of
energy.

Within IPT the self energy is calculated to second-order in $U$.  Using
the Langreth theorem~\cite{langreth} we may calculate the lesser and
greater self energies according to 
\begin{equation} \Sigma^\gtrless_{\mathrm{int}}(t) =
U^2[g^\gtrless(t)]^2g^\lessgtr(-t).
\end{equation}
The lesser and greater self energies are then Fourier transformed to
real frequency. The lesser (greater) self energy gives us the
particle (hole) spectral weight as a function of frequency for the
retarded self-energy.  Using the Keldysh Green function relation
$\Sigma^r_{\mathrm{int}}(\omega) - \Sigma^a_{\mathrm{int}}(\omega) =
\Sigma^>_{\mathrm{int}}(\omega) -\Sigma^<_{\mathrm{int}}(\omega)$, we
may express the retarded self energy in the spectral form
\begin{equation}
\label{selfenergy_form} 
\Sigma^r_{\mathrm{int}}(\omega)
= \frac{1}{2\pi }\int d\epsilon
\frac{\mathrm{Im}\left[\Sigma^<_{\mathrm{int}}(\epsilon)-
\Sigma^>_{\mathrm{int}}(\epsilon)\right]}{\omega - \epsilon + i\eta} .
\end{equation} 
The nonequilibrium DMFT self consistent equations for the
interacting and bath Green function are therefore
\begin{eqnarray}
G^r(\omega) &=& \int d\epsilon \frac{D(\epsilon)}{\omega-\epsilon-\Sigma^r_{int}(\omega)} \\
g^r(\omega)^{-1} &=& \omega+i\eta - t^2G^r(\omega) .
\end{eqnarray}
where $\rho_0(\omega) = -\pi^{-1}\mathrm{Im}[g^r(\omega)] $ is the 
new bath DOS.
For the new Keldysh bath Green functions we replace the non-interacting
DOS in Eq.~(\ref{bathGF}) with $\rho_0(\omega)$. The momentum
independence of the self energy and the translational invariance ensure
that the wave vectors remain good quantum numbers.
We iterate these equations until the bath
Green function and interacting Green function converge.

\begin{figure}[h]
\includegraphics[angle=0, width=8cm]{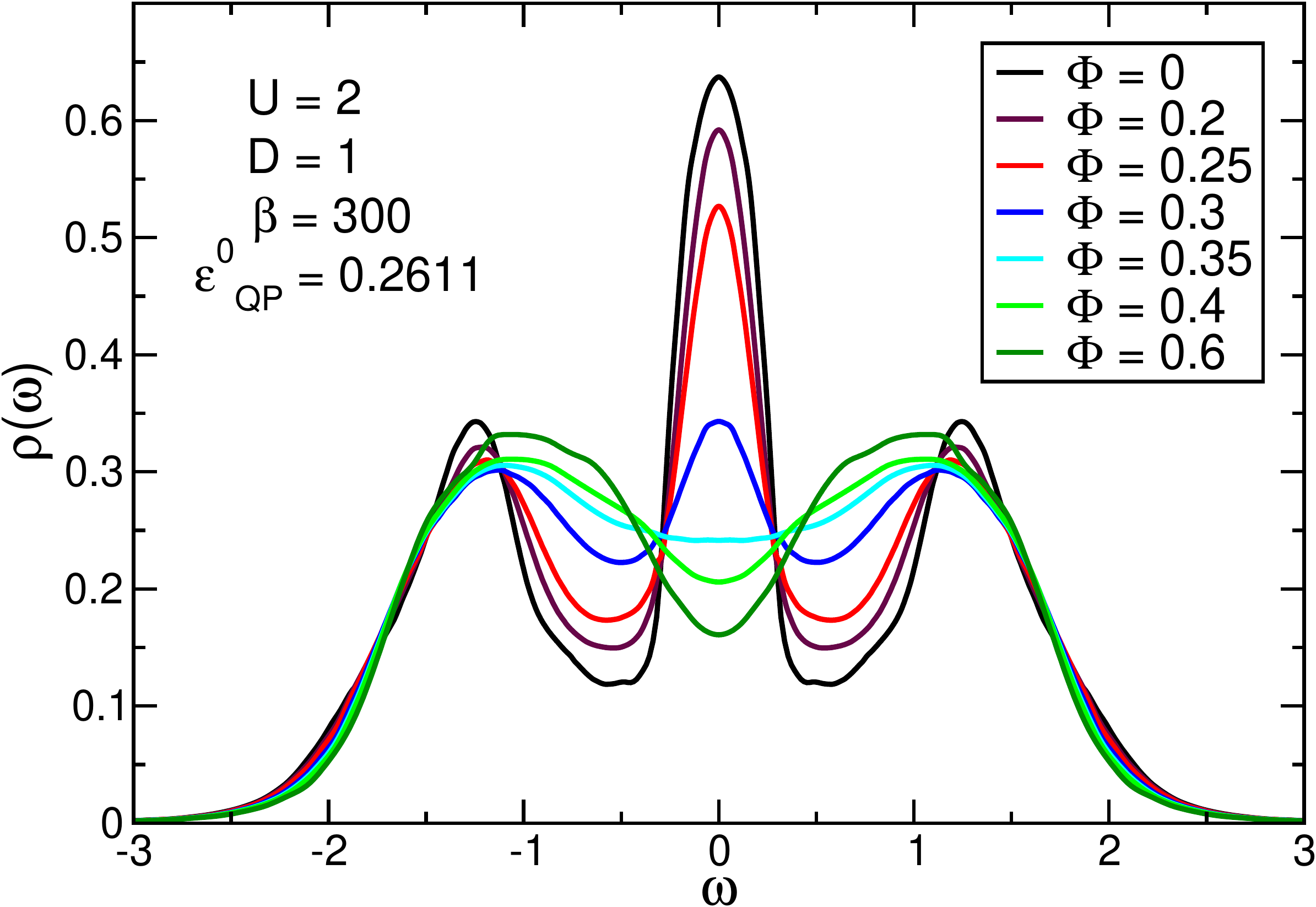}
\caption{(color online) Interacting local spectral functions for $U=2$, inverse
temperature $\beta = 300$ and half bandwidth $D=1$ plotted for a range of
chemical potential biases, $\Phi$.  At this value of the Coulomb
interaction $U$ the quasi-particle peak is destroyed continuously as
$\Phi$ is increased.}
\label{fig2}
\end{figure}

From previous works in the equilibrium system~\cite{GeorgesDMFT}, it is
known that by increasing the temperature the quasi-particles are
destroyed when the thermal fluctuations surpass a certain low-energy
scale. This low-energy scale is given by the renormalized Fermi energy
$\varepsilon_{QP}$ which we define as the half-width at half maximum
(HWHM) of the quasi-particle peak and is similar to $\epsilon_F^*=ZD$
where
$Z=[1-\partial_\omega\mathrm{Re}[\Sigma^r(\omega)]|_{\omega=0}]^{-1}$ is
the quasi-particle weight, as found in Georges
{\it et.al}~\cite{GeorgesDMFT}. As the temperature is increased at zero
bias $\Phi=0$, the quasi-particles are destroyed when
$k_BT_c\sim\varepsilon^0_{QP}$ where $T_c$ is the critical temperature.
$\varepsilon^0_{QP}$ is the quasi-particle half bandwidth at zero bias.

\begin{figure}[h]
\includegraphics[angle=0, width=8cm]{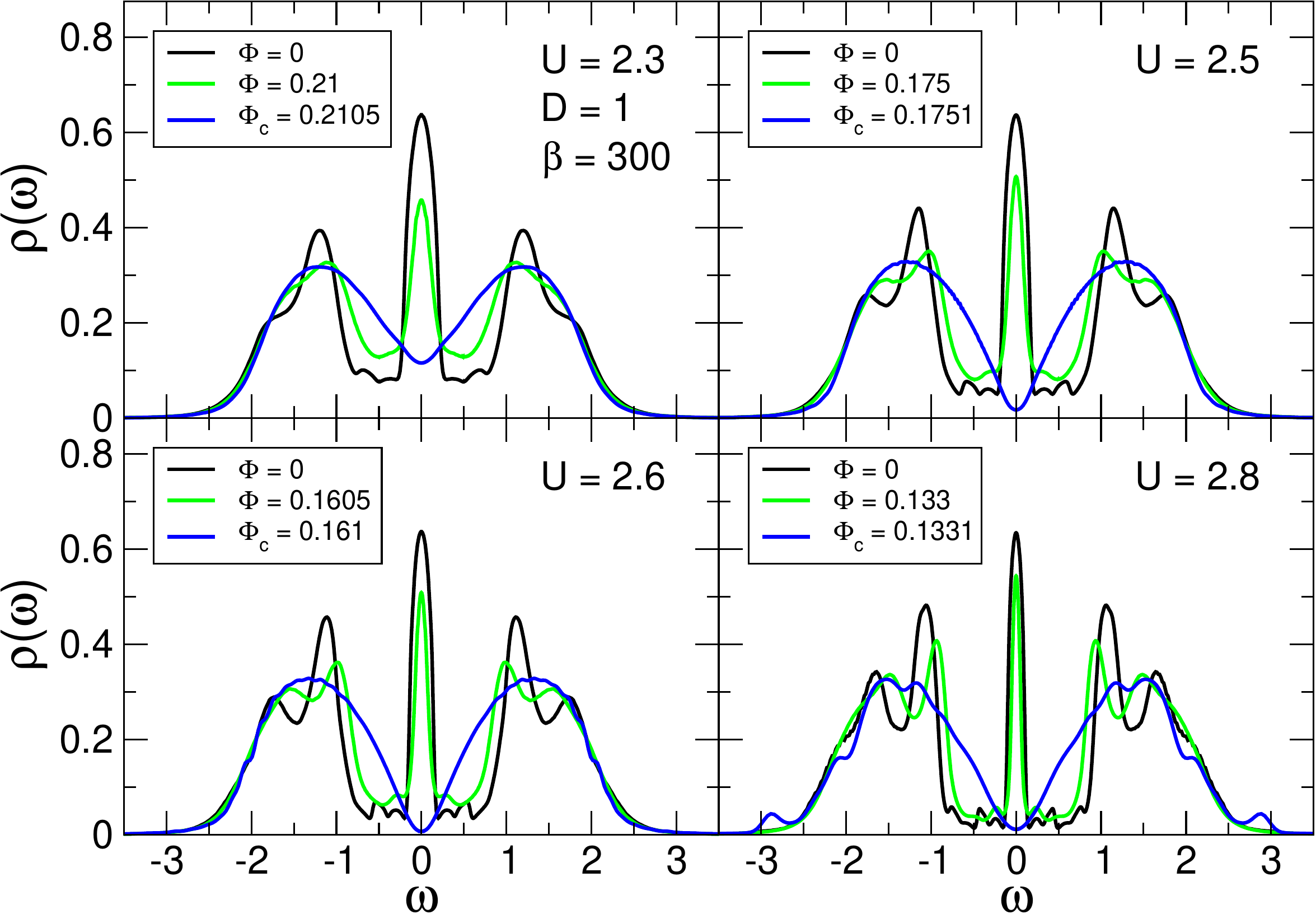}
\caption{(color online) Interacting local spectral functions for $U = 2.3, 2.5, 2.6,$
and $2.8$ at inverse temperature $\beta = 300$ and band width $D=1$.
These plots depict the sudden disappearance of the quasi-particle peak
at the critical voltage $\Phi_c$.  The spectral functions are plotted at
$\Phi=0$, at $\Phi$ just before the quasi-particle destruction, and at
$\Phi=\Phi_c$.}
\label{fig3}
\end{figure}

\begin{figure}[h]
\includegraphics[angle=0, width=8cm]{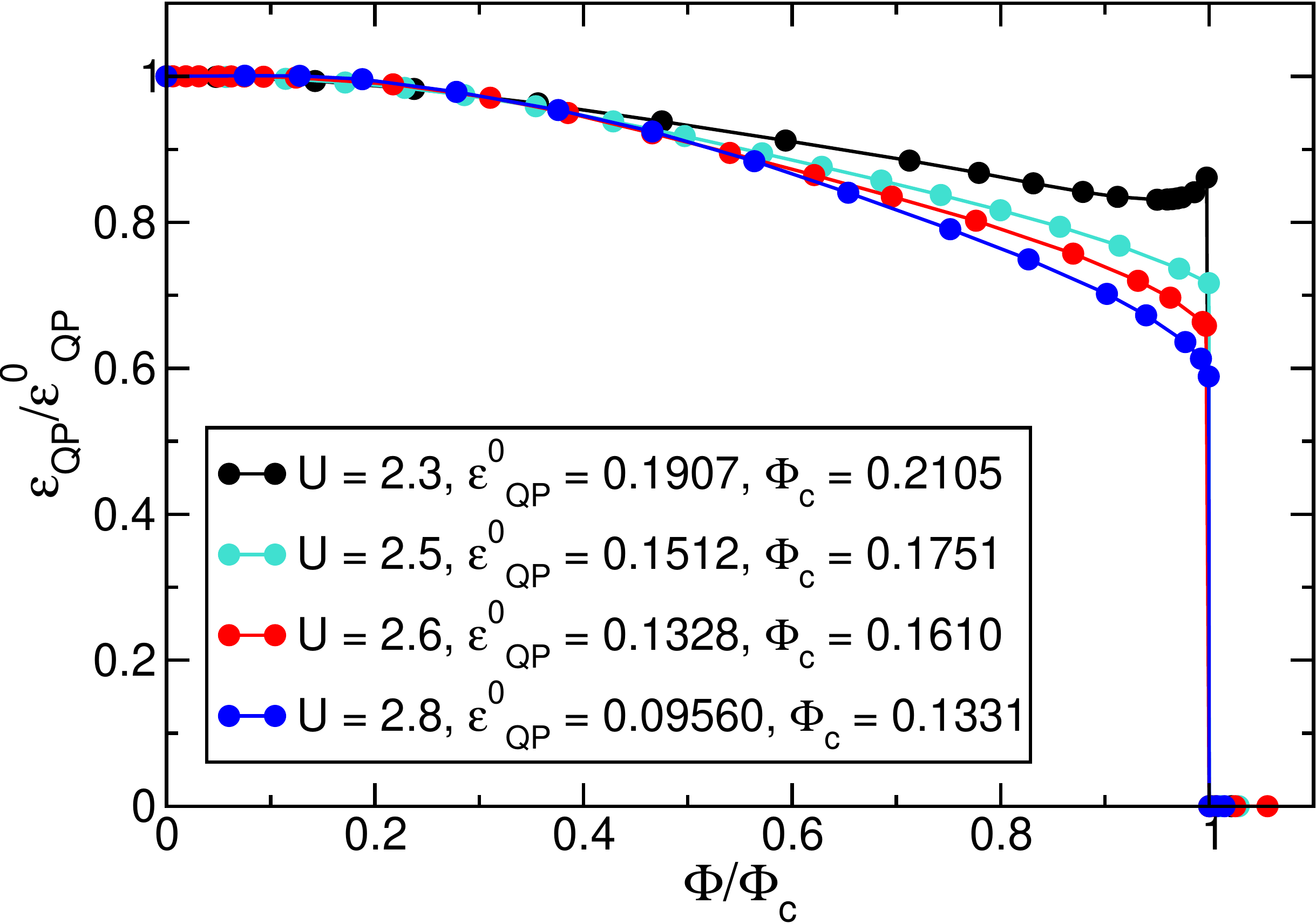}
\caption{(color online) Quasi-particle energy $\varepsilon_{QP}$ as a function of the
chemical potential bias $\Phi$  at $\beta = 300$ and $D=1$.  The
quasi-particle energy is scaled to the quasi-particle energy at zero
bias, $\varepsilon^0_{QP}$, and the chemical
potential bias is scaled to the critical bias at which the
quasi-particle peak is destroyed, $\Phi_c$.  For small bias,
$\Phi\lesssim 0.4\Phi_c$, the quasi-particle energies scale to a single curve.
In this range of values for $U$ the disappearance of the quasi-particle
peak is strongly discontinuous.}
\label{fig4}
\end{figure}

Figures~\ref{fig2}-\ref{fig4} show the evolution of quasi-particle
spectra as a function of bias $\Phi$. The destruction of the
Fermi liquid is easily understood by enhanced dephasing of particles and
holes, which is a consequence of the opening up of phase space available for
the particles/holes to scatter into through $\Phi$ and by thermal
fluctuations. Both of these effects lead to a finite lifetime for the
electron at the Fermi energy.
To lowest order in $\omega$, $T$
and $\Phi$ the imaginary part of the self energy obeys the
relation~\cite{Oguri}
\begin{equation}\label{selfexpansion}
\lim_{\omega,T,\Phi\rightarrow0} \mathrm{Im}[\Sigma^r(\omega)] 
\propto \left[\omega^2 +(\pi k_BT)^2 +\frac{3}{4}\Phi^2\right] .
\end{equation}
The interacting spectral function
$\rho(\omega)$ is calculated both as a function of the Coulomb
interaction $U$ and the applied chemical potential bias $\Phi$.  For
$U<U_d$ with $U_d\simeq 2.3D$, the destruction of the quasi-particle
peak is continuous, and exhibits similar behavior to the metal-insulator
transition in the crossover regime ($U<U_{c1}$) of equilibrium DMFT.
Our results at $U=2D (<U_d)$ are shown in Figure~\ref{fig2}.  As the
chemical potential bias is increased the quasi-particle weight smoothly
shifts towards the upper and lower Hubbard bands, and the
quasi-particle peak disappears when $\Phi\sim\varepsilon^0_{QP}$.

For $U_d<U< U_c$ the quasi-particle peak is discontinuously destroyed at
a critical chemical potential bias $\Phi_c$. In Figure~\ref{fig3} we
have plotted the interacting spectral functions for $U/D=2.3,2.5,2.6,$
and $2.8$ to illustrate the sudden disappearance of the quasi-particle
peak.  The spectral function for each value of $U$ is shown for
$\Phi=0$, $\Phi$ just before the transition, and finally at the
transition $\Phi=\Phi_c$. This discontinuous transition in nonequilibrium
by bias is reminiscent of the discontinuous transition by temperature in
the equilibrium DMFT.

In Figure~\ref{fig4} the scaled quasi-particle energy
$\varepsilon_{QP}/\varepsilon^0_{QP}$ is plotted versus the scaled
chemical potential bias $\Phi/\Phi_c$.  For $\Phi\lesssim 0.4\Phi_c$ the
curves for $U/D=2.3,2.5,2.6$ and $2.8$ scale onto a single curve.  At
$\Phi=\Phi_c$ it is clearly visible that the quasi-particle peak
disappears discontinuously for each value of $U$.

\begin{figure}[h]
\includegraphics[angle=0, width=8cm]{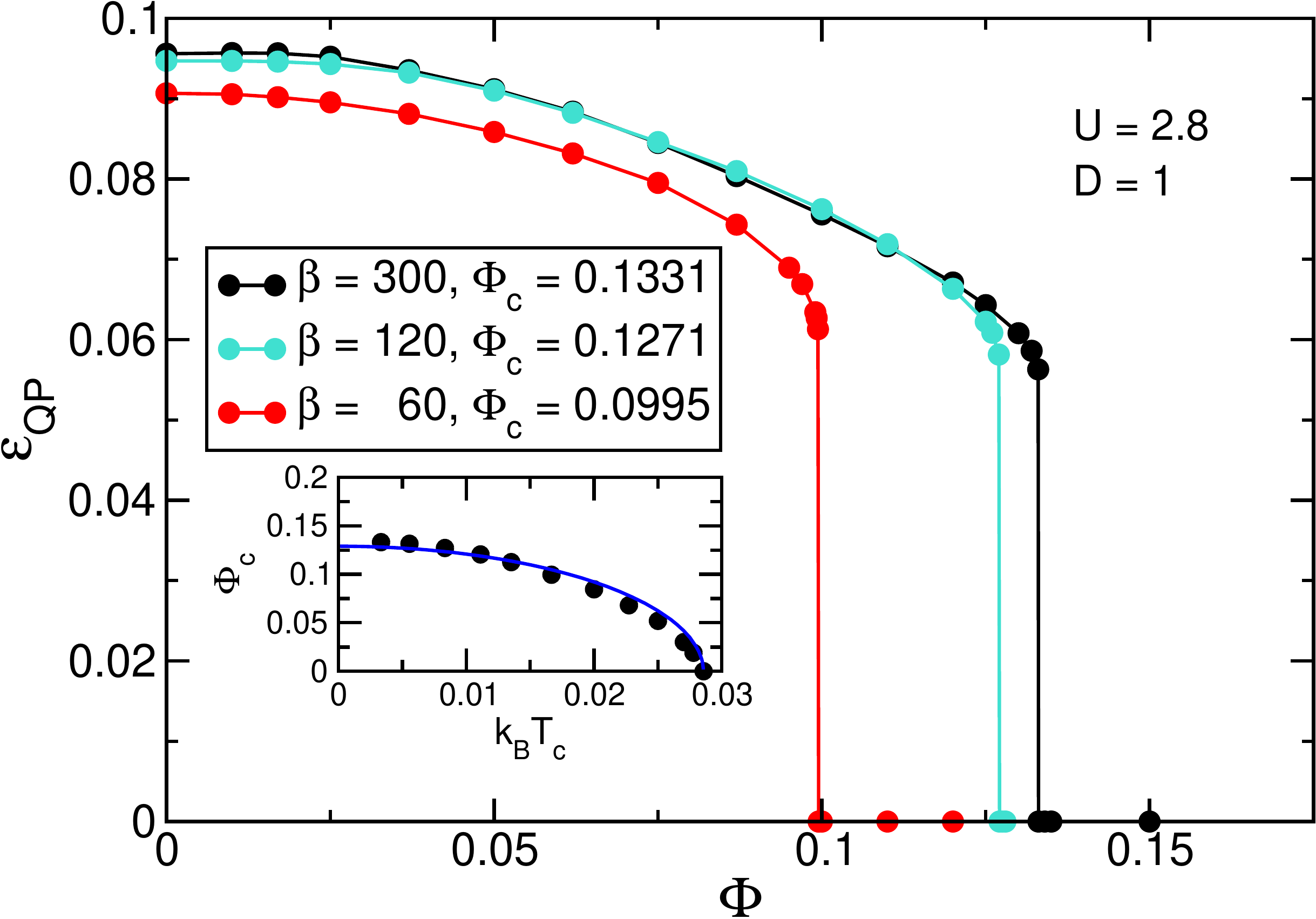}
\caption{(color online) Quasi-particle energy $\varepsilon_{QP}$ as a function of the
chemical potential bias $\Phi$  at $U = 2.8$ and $D=1$ for different
values of the inverse temperature $\beta$.  As the temperature is
increased, the disappearance of quasi-particles remains discontinuous
and the critical bias $\Phi_c$ is lowered.  The inset gives the
temperature dependence of the critical bias.  The data (black circles)
was fit (blue curve) to the function, $(\varepsilon^0_{QP})^2 \sim a(\pi
k_BT_c)^2+b\Phi_c^2$, where $a=1.1$ and $b=0.54$.}
\label{fig5}
\end{figure}

\begin{figure}
\includegraphics[angle=0, width=8cm]{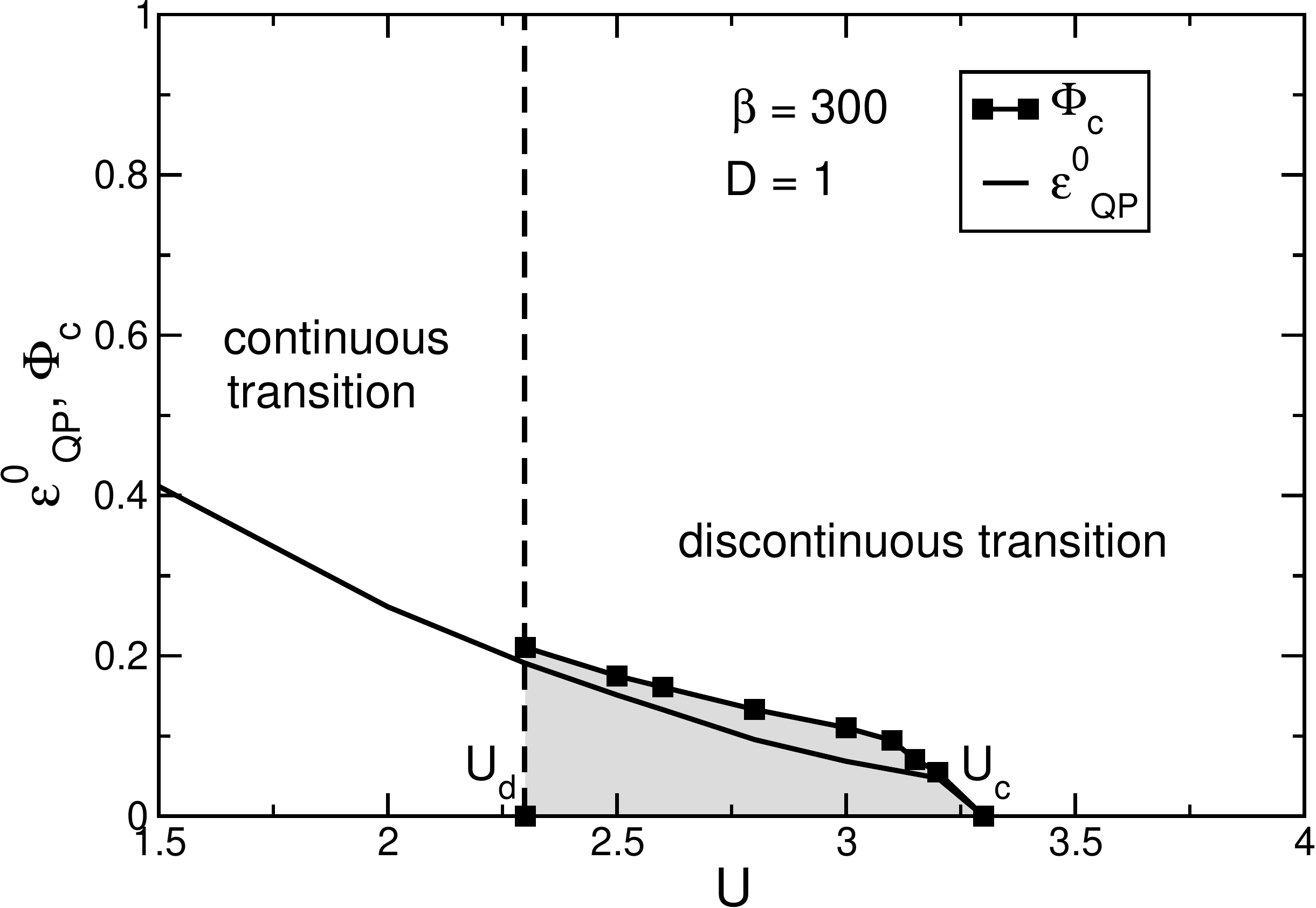}
\caption{The critical bias $\Phi_c$ is plotted as a function of the
Coulomb interaction $U$ at inverse temperature $\beta = 300$ and
bandwidth $D=1$.  We have also plotted the renormalized Fermi energy
$\varepsilon^0_{QP}$ at $\Phi = 0$.  In the region $U_d<U<U_c$ the
quasi-particle peak is destroyed discontinuously at $\Phi_c$.  When
$U<U_d$ the quasi-particle peak disappears smoothly with increasing
chemical potential bias.  At this temperature $\Phi_c\sim\varepsilon^0_{QP}$. }
\label{fig6}
\end{figure}

In Figure~\ref{fig5} we take a closer look at the temperature
dependence of the transition in the discontinuous region.  At $U=2.8$,
which is well into this region, the quasi-particle disappearance remains
discontinuous for the entire range of temperatures.  Therefore the
discontinuous nature of the transition is very robust and is not
effected by the temperature.  The inset yields the dependence of
the critical bias upon temperature.  
Motivated by the form of the imaginary part of the self energy
[Eq.~(\ref{selfexpansion})],  we
anticipate that the destruction of the quasi-particles will occur when
$(\varepsilon^0_{QP})^2 \sim a(\pi k_BT_c)^2+b\Phi_c^2$ at critical
values for the temperature $T_c$ and the bias $\Phi_c$.  In
Figure~\ref{fig5} we fit $\Phi_c$ versus $T_c$ to this function
and find that $a=1.1$ and $b=0.54$. Considering the nonlinear effects in
the numerical results, the obtained values $a$ and $b$ are in reasonable
agreement with the estimates 1 and $\frac34$ based on
Eq.~(\ref{selfexpansion}).

The critical bias $\Phi_c$ and renormalized Fermi energy
$\varepsilon^0_{QP}$ are plotted versus $U$ in Figure~\ref{fig6} at
$\beta=300$.  The onset of the discontinuous transition is marked at
$U_d\simeq2.3D$ and the Mott-insulator transition at $\Phi=0$ is given
at $U_c\simeq3.3D$.  We see that the transition occurs at
$\Phi_c\sim\varepsilon^0_{QP}$ which further justifies the function used
for fitting $\Phi_c$ versus $k_BT_c$ of Figure~\ref{fig5}.

In this work we have presented a theoretical model for nonequilibrium
dynamical mean-field theory and performed the iterative perturbation
calculations within the dynamical mean-field theory.  In a lattice where
the statistics are determined by the presence of multiple chemical
potentials, the quasi-particle properties are strongly dependent upon
the strength of the Coulomb interaction and the chemical potential bias,
$\Phi$. For $U<U_d$ with $U_d\simeq 2.3D$ for the bandwidth $D$, the
quasi-particle disappears smoothly with increasing chemical potential
bias. The disappearance of the quasi-particles is caused by the opening
up of phase space for the electrons to scatter into at the Fermi energy,
resulting in a finite lifetime for the electrons. The perturbation
theory also predicts that in the region $U_d<U<U_c$, where
$U_c\simeq3.3D$ marks the Mott-insulator transition in equilibrium, the
quasi-particle particles may be destroyed discontinuously at a critical
bias, $\Phi_c$.

This work was supported by NSF DMR-0426826 and we acknowledge the CCR
at the SUNY Buffalo for computational resources.

\end{document}